\documentclass[reqno]{article}
\usepackage{vmargin,amstex}
\setpapersize{A4}
\newcommand{\ga}{\alpha}
\newcommand{\gb}{\beta}
\newcommand{\gc}{\gamma}
\newcommand{\gC}{\Gamma}
\newcommand{\gd}{\delta}

\newcommand{\gep}{\epsilon}

\newcommand{\gf}{\phi}

\newcommand{\gk}{\kappa}
\newcommand{\gm}{\mu}

\newcommand{\gT}{\Theta}

\newcommand{\gy}{\eta}

\newcommand{\di}{\partial}
\newcommand{\bdi}{\bar{\partial}}
\newcommand{\co}{\nabla}

\newcommand{\bs}{\bar{s}}
\newcommand{\bz}{\bar{z}}

\newcommand{\bgk}{\bar{\gk}}

\newcommand{\bcF}{\bar{\cF}}

\newcommand{\cF}{\mathcal F}%
\newcommand{\lb}[1]{\label{#1}}
\newcommand{\Eq}[1]{(\ref{#1})}

\renewcommand{\[}{\begin{eqnarray}}
\renewcommand{\]}{\end{eqnarray}}
\newcommand{\nn}{\nonumber}
\newcommand{\non}{\nonumber \\ }

\begin{document}

\arraycolsep3pt

\thispagestyle{empty}
\begin{flushright} hep-th/9904097\\
                   KCL-MTH-99-12
\end{flushright}
\vspace*{2cm}
\begin{center}
 {\LARGE \sc On The Energy Momentum Tensor\\[1ex]
             of the M-Theory Fivebrane\\
 \vspace*{1cm}}
 {\sl
     Oliver B\"arwald\footnote[1]{Supported by the EC under TMR 
     contract ERBFMBICT972717,},
     Neil D.\ Lambert and
     Peter C.\ West\\
oliver, lambert, pwest@@mth.kcl.ac.uk\\
 \vspace*{6mm}
     Department of Mathematics, King's College London\\
     Strand, London WC2R 2LS, Great Britain\\
 \vspace*{6mm}}
{April 15, 1999}\\
\vspace*{1cm}
\begin{minipage}{11cm}\footnotesize
  \textbf{Abstract:} We construct the energy momentum tensor for the
  bosonic fields of the covariant formulation of the M-theory
  fivebrane within that formalism. We then obtain the energy for
  various solitonic solutions of the fivebrane equations of motion.
\end{minipage}
\end{center}
\section{Introduction}

There exist two basic extended solitonic objects in eleven dimensional
M-theory, a membrane and a fivebrane. The membrane's dynamics are
described by eight scalar fields and their spinorial superpartners and
can be derived from the usual super-p-brane action \cite{BeSeTe87}. On
the other hand the dynamics of the fivebrane are much more complicated
since they include a self-dual three form tensor field.  The equations
of motion for this system have only been obtained relatively recently
an provide us with an interesting physical system which has received
substantial attention.

There are essentially two formalisms for the dynamics of the
M-fivebrane.  The first is based on the manifestly covariant
superembedding formalism \cite{HowSez97,HoSeWe97a} which we will study
here. There is a second approach based on the non-covariant
description of \cite{PerSch97,AgPaPoSch97}. This latter approach was
then later expanded into a covariant form \cite{BLNPST97a} by the
introduction of an auxiliary scalar. In this case the equations of
motion can be obtained from an action and hence many physical
quantities such as the energy density can be readily obtained.
However, there are serious objections to the use of an action for the
M-fivebrane \cite{Wit96,LamWes98b} due to subtleties associated with the
self-duality constraint.

We are therefore interested here in extending the covariant
superembedding approach which does not invoke the use of an action.
To date a draw back has been that various physical quantities such as
the energy have not been identified within this formalism. Thus in
this paper we aim to derive directly from the covariant equations of
motion the full non-linear energy momentum tensor for bosonic fields
of the M-theory fivebrane.

In addition the covariant equations of motion can lead to lengthy
calculations and one expects that a knowledge of the energy, and more
generally the energy momentum tensor, will offer physical insights
into the structure of this highly nonlinear theory. This should also
be helpful when finding and studying new solutions. Finally, since in
the action framework the energy is known it might also lead to a
better understanding of the relation between the two approaches.

This paper is organised as follows: we start by giving a brief
introduction into the covariant formulation of the M-fivebrane. We
then construct a two parameter family of all symmetric second rank
tensors which are covariantly conserved when the fields obey their
equations of motion. After fixing the free parameters using the six
dimensional supersymmetry we are naturally led to a unique form for
the energy momentum tensor. As expected the resulting energy density
is positive definite.

Following this we will evaluate the tensor for a number of solitonic
solutions to the M-fivebrane equations of motion which involve the
self-dual three form \cite{HoLaWe98,GaLaWe98}.  We reproduce the
results for the energy in all cases where it was previously obtained
using the Hamiltonian formalism and also obtain it for one additional
case; two intersecting self-dual strings.

\section{The M-Theory Fivebrane}

Let us consider an M-fivebrane in the $x^0,x^1,x^2,...,x^5$ plane.
The field content consists of 
five scalars $X^{a'}$, $a' = 6,7,8,9,10$ and a 16 component spinor $\Theta$
corresponding to the breaking of translation and half the spacetime 
supersymmetry respectively. However it also contains an antisymmetric
second rank tensor gauge field $B$ whose field strength obeys a
self-duality condition.

The classical equations of motion of the fivebrane in the absence of
fermions and background fields are \cite{HoSeWe97a}
\[
G^{ m n} \co_m \co_n X^{a'}= 0,
\]
and
\[
G^{ m n} \co_{ m}H_{ n p q} = 0.
\]
We use $m,n,p,\ldots=0,1,\ldots,5$ and $ a, b, c,\ldots=0,1,\ldots,5$
for world and tangent indices respectively.  The symbols that occur in
the equations of motion are defined as follows: the usual induced
metric for a $p$-brane is given, in static gauge and flat background
superspace, by
\[
g_{ m n} \equiv \eta _{ m n}+ \di _{ m}X^{a'} \di _{
  n}X^{b'}\delta _{a' b'}.
\]
The covariant derivative in the equations of motion is defined using
the Levi-Civita connection associated with the metric $g_{mn}$.

We define the world surface sechsbein associated with the above metric
in the usual way via $e_m{}^a \eta_{ab} e_n{}^b\equiv g_{mn}$. There is
another inverse metric $G^{ m n}$ which occurs in the equations of
motion, mediating the coupling between the scalars and the gauge field
and
the self-coupling of the latter. It is related to the usual induced
metric given above by the equation
\[
G^{mn} \equiv (e^{-1})_c{}^m \eta^{ c a} m_{ a}^{\ d} m_{ d}
^{\ b} {(e^{-1})}_b{}^n, \lb{bigG}
\]
where the matrix $m$ is given by
\[
m_{ a}^{\ b} \equiv \delta_{ a}^{\ b} -2h_{ a c d}h^{ b c d}.
\]
The field $h_{ a b c}$ is a three-form which is
self-dual
\[
h_{ a b c}= {1\over3!}\varepsilon_{ a b c d e f}h^{ d e f},
\]
with $\varepsilon^{012345}=1$ and $\gy_{ab}={\rm
  diag}(-1,+1,\ldots,+1)$, but it is not the curl of a three-form
gauge field. It is related to the field $H_{ m n p}\equiv 3\di
_{[m}B_{np]}$ which appears in the equations of motion and is the curl
of the two-form gauge field $B_{np}$, but $H_{mnp}$ is not self-dual
in the linear sense. The relationship between the two fields is given
by
\[
H_{ m n p}= e_{ m}^{\ a} e_{ n}^{\ b} e_{ p}^{\ c} {({m }^{-1})}_{
  c}^{\ d} h_{abd}.
\]
Clearly, the self-duality of $h_{abd}$ transforms into a
condition for $H_{ m n p}$ and vice versa for the Bianchi identity
$dH=0$.

\section{Constructing the Tensor}

\subsection{The Three-Form Case}

With all scalar fields set to zero the fivebrane dynamics reduce to
the a system involving a self-interacting three-form in 6-dimensional
flat Minkowski-spacetime. For convenience we shall work in the
tangent frame in this section. In this case the self-dual
three-form tensor field $h_{abc}$ equation of motion becomes \cite{HoSeWe97b}
\[
m^{ab}\di_a h_{bcd}=0,\lb{eom}
\]
where we write $m_{ab}$ as
\[
m_{ab} = \gy_{ab} - 2 k_{ab},
\]
defining a new matrix $k_{ab}$ by
\[
k_{ab} \equiv h_a{}^{cd} h_{bcd}.
\]
We shall need some consequences of the self-duality of $h_{abc}$,
namely
\[
k^a{}_a=0,
\]
and
\[
k^{ab}k_{bc} = \frac16 \gd^a{}_c k^2,
\]
where
\[
k^2 \equiv k^{ab} k_{ab}.
\]
We now want to find the energy momentum tensor associated with
this system. To be precise we want to construct a second rank
symmetric tensor $T_{ab}$ obeying the conservation equation
\[
\di^a T_{ab}=0.
\]
Observe that the equation of motion \Eq{eom} has a symmetry, we
can send $h_{abc}$ to $-h_{abc}$ and the equation remains
unchanged. Demanding that the energy momentum tensor respects this
symmetry implies that $h_{abc}$ can appear only quadratically i.e.\ in
the form  of $k^2$ or $k_{ab}$.

The most general ansatz compatible with this restriction is
\[
T_{ab} = f_1(k^2) \gy_{ab} + f_2(k^2) k_{ab},
\]
where $f_1$ and $f_2$ are two arbitrary functions. From \Eq{eom} we can
obtain an equation of motion for $k_{ab}$, namely
\[
m^{ab} \di_a k_{bc} = \di^a k_{ac} -2 k^{ab} \di_a k_{bc}=0.
\]
We can iterate this equation as follows
\[
\begin{array}{rcl}
\di^a k_{ab} &=& 2 k^{ac} \di_a k_{cb},\\[9pt]
&=& 2 \di_a (k^{ac}k_{cb}) - 2  k_{cb} \di_a k^{ac},\\[9pt]
&=& \tfrac13 \di_b k^2 -4 k_{cb} k^{ad} \di_a k_d{}^c,\\[9pt]
&=& \tfrac13 \di_b k^2 +4 k_{cb} k_d{}^c \di_a  k^{ad} -4  k_{cb}\di_a
k^{ad} k_d{}^c,\\[9pt]
&=& \tfrac13 \di_b k^2 +\tfrac23 k^2 \di^a  k_{ab} - \tfrac23
k_{ab}\di^a k^2,\\[9pt]
&=& \tfrac13 m_{ab} \di^a k^2 + \tfrac23 k^2 \di^a k_{ab}.
\end{array}
\]
Rewriting the last line gives 
\[
\di^a k_{ab} = \frac{\tfrac13 m_{ab}\di^a k^2}{1-\tfrac23 k^2}.
\]
Plugging this into our ansatz we get
\[
\di^a T_{ab} &=&\di_b f_1(k^2) + k_{ab} \di^a f_2(k^2) + f_2(k^2) \di^a
k_{ab},\non
&=& f_1'\di_b k^2 + k_{ab} f'_2 \di^a k^2+f_2\frac{m_{ab}\di^a k^2}{3- 2
  k^2},\\
&=& \left(f'_1 + f_2 \frac1{3-2k^2}\right) \di_b k^2 + \left(f'_2 -2f_2
  \frac1{3-2k^2}
\right)
k_{ab}\di^a k^2.\nn
\]
Demanding conservation implies that the expressions in the two
brackets should vanish. This gives two ordinary differential equations for
$f_1$ and $f_2$. The general solutions are
\[
f_2 = \frac\ga{3-2k^2},\qquad \mbox{and} \qquad
f_1 = -\frac12 \frac\ga{3-2k^2} + \gb,
\]
with two constants $\ga$ and $\gb$.
Hence the most general conserved symmetric tensor built out of
$k_{ab}$ has the form
\[
T_{ab} &=& \gb \gy_{ab} -\frac12 \frac\ga{3-2k^2} \gy_{ab} + 
\frac\ga{3-2k^2} k_{ab}
= \gb \gy_{ab} - \frac12 \frac\ga{3-2k^2} m_{ab}.
\]
\subsection{The Covariant Generalisation}

There is an obvious generalisation of this tensor to the case of
active scalars, namely by interpreting the flat-space coordinates of
the previous section as coordinates of the tangent-frame. However it
is not obvious that this doesn't spoil our earlier reasoning. The
setup remains the same apart from the equation of motion \Eq{eom}
which becomes \cite{HoSeWe97a}
\[
m^{ab} \co_a h_{bcd} =0.
\]
The main difference between working in flat space and working in the
moving frame is the fact that the covariant derivatives do not
commute. But since all our calculations only involve a single
derivative everything goes through as before and we find that
\[
T_{ab} = \gb \gy_{ab} - \frac12 \frac\ga{3-2k^2} m_{ab},
\]
is covariantly conserved
\[
\co^a T_{ab} =0.
\]
We can now find the canonical tensor simply by switching to the coordinate
frame using the sechsbeins. We will furthermore include a conventional
factor of $\sqrt{-g}$ into our tensor to make sure that
\[
E_{\rm tot }=-\int d^5\!x T^{00},
\]
is the invariant total energy. This allows us to interpret the
fivebrane-model alternatively as a nonlinear theory in six-dimensional
flat space where $d^6\!x$ rather than $\sqrt{-g}d^6\!x$ is the
natural measure.

Rescaling $\ga$ for later convenience
we get
\[
T^{mn} = \gb\sqrt{-g}g^{mn} + \ga\sqrt{-g}Q^{-1}m^{mn},\lb{genTensor}
\]
where we introduced $Q\equiv 1-\frac23 k^2$.

\subsection{Fixing the Parameters}

The alert reader might be surprised about the appearance of two free
parameters in the conserved tensor. This is however to be expected
since the energy contains two arbitrary parameters, namely the
scale and the origin. Indeed the second constant $\gb$
multiplies the metric $g_{mn}$ which is always covariantly constant
and therefore by our construction $\gb$ is entirely arbitrary.

On the other hand in a supersymmetric theory there is an alternative
way of computing energy and momentum; they appear on the right hand
side of the anti-commutator of two supersymmetry transformations. This
relation will allow us to fix one of the parameters.

The most general form of the (2,0) supersymmetry algebra in six
dimensions is
\[
\{Q_\ga^i,Q_\gb^j\} = \gy^{ij} \gc^m_{\ga\gb} P_m + \gc^m_{\ga\gb}
Z^{ij}_m + \gc^{mnp}_{\ga\gb} Z^{ij}_{mnp},\lb{GenSuSy}
\]
where $\gy^{ij}$ is the Spin(5) invariant tensor, $P_m$ is the
momentum and $Z^{ij}_m$ and $Z^{ij}_{mnp}$ are central charges. The
spinor indices $\ga,\gb,\ldots$ run from 1 to 4 as do the internal
Spin(5) indices $i,j,\ldots$. The $\gc$-matrices should not be
confused with ordinary $\gC$-matrices satisfying $\{\gC^m,\gC^n\}=2
\gy^{mn}$.  They arise as building blocks of the eleven dimensional
$\gC$-matrices via
\[
\gC^m = \left(
\begin{array}{cc}0 & \gc^m \\ \tilde{\gc}^m & 0
\end{array}
\right).
\]
The basic relations are
\[
\{ \gc^m, \gc^n \} \equiv \gc^m \tilde{\gc}^n + \gc^n \tilde{\gc}^m=2\gy^{mn},
\]
with $\tilde{\gc}^m = \gc^m$ for $m\neq0$ and $-\tilde{\gc}^0 =
\gc^0=1$. The antisymmetric product is defined as
\[
\gc^{m_1m_2m_3\ldots} \equiv \gc^{[m_1}\tilde{\gc}^{m_2} \gc^{m_3} \ldots,
\]
and one also has the following duality relation
\[
\gc^{m_1m_2\ldots m_n}= - \frac1{(6-n)!}(-1)^{\frac{n(n+1)}{2}}
\gep^{m_1m_2\ldots m_n m_{n+1}\ldots m_6}\gc_{ m_{n+1}\ldots m_6}.
\]

We need the local version of equation \Eq{GenSuSy}. Recall that to every
symmetry of a physical system one can construct an associated
conserved current $j_\gm$ which upon integration gives rise to a time
independent charge
\[
Q = \int d^{d-1}\!x j_0,
\]
which acts as a generator of the symmetry on the fields of the model via
\[
\gd \gf = [Q,\gf].
\]
The expression on the right hand side of the last equation is
understood to be evaluated in terms of the Poisson brackets of the
fundamental fields.

Hence the local version of equation \Eq{GenSuSy} is given by
\[
\int d^5\!x \{Q_\ga^i,j_{0\gb}^j\}=\int d^5\!x\gy^{ij} \gc^m_{\ga\gb}
T_{0m}\mbox{ + central charge terms,}
\]
where $j_{m\gb}^j$ is the supercurrent and $T_{mn}$ the energy
momentum tensor. We can rewrite this as
\[
\int d^5\!x \gd_\ga^i j_{0\gb}^j=\int d^5\!x\gy^{ij} \gc^m_{\ga\gb}
T_{0m}\mbox{ + central charge terms,}
\]
from which we learn that the energy momentum tensor can be obtained as
the supervariation of the supercurrent.

To make our calculation simpler we shall consider the linearised
supersymmetry and also ignore all interactions. This will allow us to
use the free-field Poisson brackets and yet will still determine 
the constant $\gb$.

Since we are only interested in the bosonic part of the energy
we can focus on the part of the supercurrent which is
of the form
\[
j=\gT B\mbox{ + terms cubic in $\gT$},
\]
where $B$ is constructed out of bosonic fields only. The
full supervariation of the supercurrent is then given by 
\[
\gd j = B \gd \gT,
\]
plus terms which vanish if we set the Fermions to zero.
To determine $B$ consider the supervariation of $\bar{\gT}$, the
Dirac conjugate of $\gT$. We have
\[
\gd\bar{\gT}  = \{Q,\bar{\gT}\} = \int d^5\!x \{j,\bar{\gT}\}
              = \int d^5\!x \{\gT B,\bar{\gT}\}
              = \int d^5\!x B \{\gT,\bar{\gT}\}= B,
\]
where we used the free Fermion Poisson bracket
$\{\gT(x),\bar{\gT}(x')\}=\gd(x-x')$ and ignored terms which vanish if
we set the Fermions to zero. Hence can read off all information from
the supervariation of the Fermions, which was calculated for the most
general case in \cite{GaLaWe98}. We have, at linearised level,
\[
\gd \gT_\gb^j = \frac12\gep^{\ga i}\di_m X^{c'} \gc^m_{\ga\gb} \gc_{c'
  i}{}^j
                -\gep^{\ga j}\frac16 h_{m_1m_2m_3} \gc^{m_1m_2m_3}_{\ga\gb}.
\]
For the supervariation of the supercurrent we find from our earlier
reasoning
\[
\gd j  &=&\frac14(\gep_1)^{\ga i}\di_m
X^{c'} (\gc^m\gc^0)_{\ga\gb} \gc_{c' i}{}^j \di_n X^{d'}
(\gc^n)^\gb{}_\gc \gc_{d'
  j}{}^k (\gep_2)_{\gc k}\non
&&-\frac1{36} (\gep_1)^{\ga i} h_{m_1m_2m_3} (\gc^{m_1m_2m_3}\gc^0)_{\ga\gb} 
 h_{m_4m_5m_6} (\gc^{m_4m_5m_6})^{\gb\gc} (\gep_2)_{\gc i}\non
&&-\frac16 (\gep_1)^{\ga i}\di_m
X^{c'} (\gc^m\gc^0)_{\ga\gb} \gc_{c' i}{}^j h_{m_1m_2m_3} (\gc^{m_1m_2m_3})^{\gb\gc}
(\gep_2)_{\gc j}\non
&&+\frac16 (\gep_1)^{\ga i} h_{m_1m_2m_3} (\gc^{m_1m_2m_3}\gc^0)_{\ga\gb} \di_m
X^{c'} (\gc^m)^{\gb\gc} \gc_{c' i}{}^j
(\gep_2)_{\gc j},
\\
&=&
\frac14(\gep_1)^{\ga i} (\gc^m\gc^0\gc^n)_\ga{}^\gc
(\gc_{c'}\gc_{d'})_i{}^k
\di_m X^{c'}\di_n X^{d'}(\gep_2)_{\gc k}\non
&&-\frac1{36} (\gep_1)^{\ga i}  (\gc^{m_1m_2m_3}\gc^0\gc^{m_4m_5m_6})_\ga{}^\gc
h_{m_1m_2m_3} h_{m_4m_5m_6} (\gep_2)_{\gc i}
\non
&&+\frac16 (\gep_1)^{\ga i}  ([\gc^{m_1m_2m_3}\gc^0,\gc^m])_\ga{}^\gc
\gc_{c' i}{}^j \di_m
X^{c'} h_{m_1m_2m_3}
(\gep_2)_{\gc j}.\nn
\]
For simplicity we take only one of the scalars fields to be active,
namely $X\equiv X^{5'}$. Using the identities
\[
h_{m_1m_2m_3}h_{m_4m_5m_6} \gc^{m_1m_2m_3}\gc^0\gc^{m_4m_5m_6}=
2 (3!)^2 \gc_m k^{m0},
\]
and
\[
\di_m X \di_n X \gc^m \gc^0 \gc^m = 2 \gc_m (\tfrac12 \gy^{m0} \di^p X \di_p X
-\di^m X \di^0 X),
\]
and neglecting the last term which is a central
charge contribution we find
\[
\gd j &=& \frac12 \gep_1 \gc_m \gep_2 \left\{
\frac12 \gy^{m0} \di^p X \di_pX - \di^mX \di^0X -4 k^{m0}\right\}.
\lb{SuSyTensor}
\]
Expanding \Eq{genTensor} up to terms quadratic in fields gives
\[
T^{mn} = (\ga+\gb)(\gy^{mn} + \tfrac12 \gy^{mn}\di^p X \di_p X -
\di^mX \di^n X) - 2 \ga k^{mn}.\lb{BraneTensor}
\]
The two expressions look similar except for the first term in
\Eq{BraneTensor} which does not appear in \Eq{SuSyTensor}.
This is to be expected,
however, since a configuration with all fields set to zero corresponds
to the vacuum and hence must have zero energy in a supersymmetric
theory. From the 11-dimensional viewpoint the same configuration is a
flat, static brane which has the constant energy density normalised to
one.

Comparing the other coefficients gives the relation $\gb=-\tfrac12
\ga$ and we find that
\[
T^{mn} = \ga \tfrac12 \frac{\sqrt{-g}}{Q} \left((2-Q) g^{mn} - 4
  k^{mn}\right),\lb{emtensor}
\]
is the unique conserved, symmetric rank two tensor that is compatible
with supersymmetry, justifying the name energy momentum tensor.
Following our earlier reasoning we can normalise the tensor by
demanding that it reduces to $\gy^{mn}$ if all fields are set to zero.
This gives $\ga=2$, a choice we adopt from now on.
For a static configuration we find the following simple formula for
the energy density
\[
E = \frac{\sqrt{-g}}{Q}\left(2-Q + 4 k^{00}\right).
\]
Recall that the field $h_{mnp}$  is not closed and the physics 
is most naturally described by $H_{mnp}$. 
We would therefore like to find 
$T^{mn}$ in terms of $H_{mnp}$. To this end we note two 
identities which can be readily derived \cite{GaLaWe98}
\[
Q &=& -{3\over H^2}\left(1 - \sqrt{1 +{2\over3}H^2}\right),\\
h_{mnp} &=& QH^{(+)}_{mnp} ,
\]
where $H^{(+)}_{mnp}$ is the self-dual part of $H_{mnp}$. Therefore
one finds
\[
k^{mn} = Q^2 H^{(+)mpq}H^{(+)n}{}_{pq}.
\]

Finally we note that we can rewrite this tensor in a much simpler form in 
terms of the natural metric $G^{mn}$ occurring in the superembedding formalism.
To this end recall the definition of the inverse metric \Eq{bigG},
\[
G^{mn} = (1+ \tfrac23 k^2)g^{mn} - 4 k^{mn} =  (2-Q) g^{mn} -4 k^{mn}.
\]
Using this we find that
\[
T^{mn} = \sqrt{-g} Q^{-1} G^{mn}.
\]
We can replace the determinant as well. Using
\[
\sqrt{-G} = Q^{-3} \sqrt{-g}
\]
we find a third expression for the tensor, namely
\[
T^{mn} = \sqrt{-G} Q^2 G^{mn}.
\]
From this final expression it is obvious that the energy, given by $E=
-T^{00}$, is always positive definite. Note that $\sqrt{-G}$ and $Q^2$
are always positive and $G^{00}$ is the time-time-component of the
metric which occurs naturally in the embedding formalism. Using our
conventions this implies that $G^{00}$ is negative definite and hence
the energy is positive definite. We note that this agrees with the
energy momentum tensor obtained using the action formulation
\cite{BLNPST97b}.

\section{Applications}

To make contact with previous work on the energy of fivebrane
configurations we now evaluate our tensor for some of the known solitonic
solutions to the fivebrane equations of motion. For most cases
expressions for the energy are also known \cite{GaGoTo98,GaLaWe98}
from the  noncovariant Hamiltonian formalism
\cite{PerSch97,AgPaPoSch97,BLNPST97a}.  Here we are able to reproduce
these known results and also to determine the energy for the intersecting
self-dual string solution of \cite{GaLaWe98,LamWes98c}.

\subsection{The Self-Dual String}
We are looking for a string-soliton whose world sheet lies in the
$(x^0,x^5)$-plane and hence take all fields to be independent of these
two coordinates. We shall denote indices ranging from
0 to 5 from indices ranging from 1 to 4 by putting a hat on the
former. We use the following ansatz \cite{HoLaWe98}
\[
X^{6'} &\equiv& \gf, \nn\\
h_{05a} &\equiv& v_a, \\
h_{abc} &=& \gep_{abcd} v^d,\nn
\]
the last equation being a consequence of the self-duality of $h$.
All other scalars and all other components of
$h_{\hat{a}\hat{b}\hat{c}}$ are set to zero. To evaluate the
  energy-momentum tensor we have to calculate the $m$-matrix. In the
  tangent frame we get:
\[
m_{\hat{a}}{}^{\hat{b}} = \left(
\begin{array}{ccc}
1+4v^2 & 0      & 0 \\
0      & (1-4v^2) \gd_a{}^b + 8 v_a v^b  & 0 \\
0      & 0      & 1+4v^2 
\end{array}\right).
\]
The usual induced metric reduces in static gauge to
\[
g_{\hat{m}\hat{n}}= \left(
\begin{array}{ccc}
-1 & 0  & 0 \\
0  & \gd_{mn} + \di_m \gf \di_n \gf  & 0 \\
0  & 0  & 1 
\end{array}\right).
\]
We shall also need the inverse metric and the associated sechsbein,
they take the form of
\[
g^{\hat{m}\hat{n}}= \left(
\begin{array}{ccc}
-1 & 0  & 0 \\
0  & \gd_{mn} + \frac{\di_m \gf \di_n \gf}{g}  & 0 \\
0  & 0  & 1
\end{array}\right),
\]
and
\[
e_{\hat{m}}{}^{\hat{a}} = \left(
\begin{array}{ccc}
1  & 0  & 0 \\
0  & \gd_m^a +c \di_m \gf \di^a \gf & 0 \\
0  & 0  & 1
\end{array}\right),
\]
where $ -g=-\det g = 1 + |\di \gf |^2$ and
$c = \frac{-1 \pm \sqrt{-g}}{|\di \gf |^2}$.

We are only interested in the energy of this
configuration. We find 
\[
-E = T^{00} = \sqrt{-g}g^{00} e_0{}^0 e^0{}_0 (2 m_0{}^0Q^{-1} -
\gd_0{}^0)= -2\sqrt{-g} (1+4v^2)Q^{-1} +\sqrt{-g}.
\]
Using $Q=1 - \tfrac23 k^2 = 1 - 16 v^4$ this reduces to
\[
E = \sqrt{-g}\frac2{1-4v^2} - \sqrt{-g}.
\]
Demanding that the solution preserves half the supersymmetries leads
to the Bogomol'nyi condition
\[
v_a = \frac12 \frac{\di_a \gf}{1+\sqrt{1+|\di \gf |^2}}.\lb{StringBogo}
\]
A bit of algebra gives $1-4 v^2= \tfrac2{1+ \sqrt{-g}}$
and hence finally
\[
E=-g = 1+ |\di \gf|^2,
\]
which agrees with the result obtained using the non-covariant formalism
\cite{GaGoTo98,GaLaWe98}.

\subsection{Neutral Strings: The Instanton}
This solution is obtained by setting all scalars to zero, and taking
the remaining fields to be independent of the $x^0$ and $x^5$
directions. Since with all scalars inactive the induced metric is the
flat metric we shall work in the tangent frame in
this section.  Furthermore we take the three-form to be
\[
h_{0ab} = \pm h_{5ab} \equiv F_{ab},
\]
with all other components set to zero. Depending on the sign the
two-form $F_{ab}$ is taken to be either self-dual or anti-self-dual.
We get the following expression for $k_{\hat a}{}^{\hat b}$,
\[
k_{\hat a}{}^{\hat b} = \left(
\begin{array}{ccc}
-F^2  & 0  & F^2 \\
0    & 0  & 0   \\
-F^2    & 0  & F^2
\end{array}\right),
\]
where $F^2=F^{ab}F_{ab}$ and the scalar $k^2$ vanishes. Focusing on the energy we have
\[
E=-T^{00}=1+4 F^2.
\]
This also agrees, up to a rescaling of $F \rightarrow \tfrac14 F$, with
the result obtained from the non-covariant formalism \cite{GaLaWe98}.

\subsection{Combining Neutral and Self-Dual Strings}

If we superpose the solutions of the two previous subsections we get
the following expression \cite{GaLaWe98} for $k_{\hat a}{}^{\hat b }$
\[
k_{\hat a}{}^{\hat b} = \left(
\begin{array}{ccc}
-F^2-2 v^2  & \sqrt2v_c F^{bc}  & F^2 \\
4 \sqrt2 v^c F_{ac}    & 2v^2\gd_a{}^b -4 v_a v^b  & -4 \sqrt2 v^c F_{ac} \\
-F^2    & \sqrt2 v_c F^{bc} & F^2-2v^2
\end{array}\right).
\]
Using the Bogomol'nyi condition \Eq{StringBogo} gives after a
lengthy calculation the following expression for the energy
\[
E = 1+ |\di \gf |^2 + (1+\sqrt{1+|\di \gf |^2})^2F^2.
\]
This result uses the unphysical field $F_{ab}$ which is not the curl
of a one-form gauge field. Recall that in general the self-dual
three-form $h_{\hat a \hat b \hat c}$ is not equal to the physical
three-form $H_{\hat a \hat b \hat c}$. The relation between $F_{ab}$
and the physical field $K_{ab}=H_{0ab}$ is given by
\[
K_{ab} &=& (1-4v^2)^{-1} F_{ab} = \tfrac12 (1+ \sqrt{1+|\di \gf |^2}) F_{ab}.
\]
Hence in terms of $K$ the energy is given by 
\[
E = 1+ |\di \gf |^2 + 4 K^2,
\]
which up to a rescaling $K \rightarrow \tfrac14 K$ agrees with the expression for the
energy obtained using the Hamiltonian formalism \cite{GaLaWe98}.

\subsection{Intersecting Self-Dual Strings: The Monopole}
As a final example we now consider two intersecting self-dual strings 
\cite{GaLaWe98,LamWes98c}. This soliton can be related  to monopole
configurations in $N=2$ supersymmetric Yang-Mills theories \cite{LamWes98c}
and the evaluation of the energy is therefore of some interest.

We have two active scalars, $X^{6'}$ and $X^{7'}$, depending on the 4
spacetime coordinates $x^0,\ldots,x^3$ and two additional
coordinates $x^4$ and $x^5$. It will be useful to introduce complex
coordinates for these
\[
z\equiv x^4+ix^5,
\]
and also to combine the two real scalar fields into a single complex scalar
field
\[
s \equiv X^{6'}+i X^{7'}.
\]
We denote the associated derivatives by
\[
\di \equiv \di_4+i\di_5.
\]
In this section we use indices $a,b,c,\ldots=0,1,2,3$ in the
tangent-frame, where we will perform all calculations, indices
$m,n,p,\ldots=0,1,2,3$ in the world-frame and put hats on these to
denote indices taking the full range from 0 to 5. In the tangent frame
we shall also use $i,j,k,\ldots =1,2,3$ to denote purely spatial indices.

In the complex coordinate system the flat metric and its inverse are given by 
\[
\gy_{\hat{a}\hat{b}} \equiv \left(
\begin{array}{cc}
\gy_{ab} & 0  \\
0  & \smallmatrix
0 & \tfrac12 \\
\tfrac12 &0
\endsmallmatrix
\end{array}\right),
\quad
\mbox{and}
\quad
\gy^{\hat{a}\hat{b}} \equiv \left(
\begin{array}{cc}
\gy^{ab} & 0  \\
0  & \smallmatrix 
0 & 2 \\
2 &0
\endsmallmatrix
\end{array}\right),
\]
where $\gy_{ab}$ and $\gy^{ab}$ denote the usual four-dimensional flat
Minkowski-metric and its inverse. In the world-frame we find
\[
g_{\hat{m}\hat{n}} = \gy_{\hat{m}\hat{n}} + \tfrac12 \di_{\hat{n}}s\di_{\hat{m}}\bs
      + \tfrac12 \di_{\hat{n}}\bs\di_{\hat{m}}s.
\]
We also need the determinant which is given by
\[
\begin{aligned}
-g= -\det g &= (1+ |\di s|^2 - |\bdi s|^2)^2 + 4 |\bdi s|^2\\
&\hphantom{=(+}+ |\di_i s|^2(1 + |\di s|^2 + |\bdi s|^2)
+ (\di_i s)^2 \bdi s\bdi \bs +  (\di_i \bs)^2 \di s\di \bs.
\end{aligned}
\lb{detg}
\]
We will only consider static solutions, i.e.\ have set $\di_0\equiv0$ and
also have expanded all expressions up to second order in the spatial derivatives
for simplicity. The process of solving the fivebrane equations starts
with making an ansatz for the six-dimensional three-form. We decompose
$h$ into four-dimensional two-forms and vectors as follows
\[
h_{abz} = \gk\cF_{ab}, \quad h_{ab\bz}=\bgk\bcF_{ab}, \quad h_{az\bz} = i v_a.
\]
Self-duality of $h$ implies that $h_{abc}= 2 \gep_{abcd} v^d$ and
$\cF_{ab} = \tfrac{i}2 \gep_{abcd} \cF^{cd}$.

Demanding preservation of half of the supersymmetry leads to the
following set of Bogomol'nyi-conditions \cite{GaLaWe98,LamWes98c}. Given in the
tangent frame they are
\[
\gk \cF_{0i} 
&=& {1\over8}\eta\left({1+|\di s|^2-|\bdi s|^2\over X^2 - |\bdi s|^2}\right)
\left(
{X^2\di_is + \di\bs\di s\di_i\bs\over X \det e}
\right)\ ,\nn\\
v_0 &=& +{i\over16}\eta\left({1+|\di s|^2-|\bdi s|^2\over (X^2-|\bdi s|^2)^2}\right)\left[
(1+|\di s|^2+|\bdi s|^2){\bar\di s\di_is\di^i\bs\over(\det e)^2}\right. \nn\\
&&\left. \ \ \ +|\bdi s|^2
{(\di s\di_i\bs\di^i\bs-\bar\di\bs\di_i s\di^i s)
\over (\det e)^2}\right]+ {i\over4}\eta{\bar\di s\over X^2 - |\bdi s|^2},\\
v_i &=& {1\over16}\eta\bar\di s\left({1+|\di s|^2-|\bdi s|^2\over (X^2-|\bdi s|^2)^2}\right)
{\epsilon_{ijk}\di^js\di^k\bs\over \det e}\ ,\nn\\
{\bar \di}s &=& -\di\bs\nn,
\]
where $\gy = \pm 1$ and the conditions for the remaining components of
$\cF$ are obtained by using its self-duality. We have used the following
convenient expressions
\[
\det e &\equiv& \sqrt{(1+|\di s|^2-|\bdi s|^2)^2+4|\bdi s|},\\
X^2    &\equiv& \frac12 (1+|\di s|^2+|\bdi s|^2+\det e).
\]
Here $\det e$ denotes not the full determinant of the vielbein but only
the part without spatial derivatives.

In the static case we have the following formula for the energy
\[
E = \sqrt{-g} \frac{2-Q-4 k_0^0}{Q}.
\]
We only need to know $Q$ and $k_0{}^0$ which are given by
\[
Q &=& 1-256v_0^2(v_0^2-2|\gk|^2\cF_{0i}\bcF^{0i}),\\
k_0{}^0 &=& -8 v_0^2 + 8 |\gk|^2\cF_{0i}\bcF^{0i}.
\]
Despite the complexity of the Bogomol'nyi conditions we finally get a
remarkably simple answer for the energy, namely
\[
E = \frac{-g}{1 + |\di s|^2 - |\bdi s| ^2},
\]
where the determinant of the spacetime metric is given by
\Eq{detg}.

If we take $s$ to be
a holomorphic function of $z$ the energy reduces to 
\[
E= 1 + |\di s|^2 + |\di_i s|^2.
\]
Reverting to the real scalar fields and setting $\di\equiv0$ we find
\[
E = 1+ (\di_i X^6)^2 + (\di_i X^7)^2,
\]
which agrees with the energy obtained using the Hamiltonian formalism
as given in \cite{GaLaWe98}.

\section{Conclusions}

In this paper we have shown that the tensor
\[
T^{mn} = \frac{\sqrt{-g}}{Q} \left((2-Q) g^{mn} - 4 Q^2 H^{(+)mpq}H^{(+)n}{}_{pq}
\right),
\]
where $Q = -\tfrac{3}{H^2}(1 - \sqrt{1 +{2\over3}H^2})$ and
$H^{(+)}$ denotes the selfdual part of $H$,
is covariantly conserved, compatible with supersymmetry and gives
expressions for the energy of solitonic configurations which agree
with the Hamiltonian expressions in all known cases.

In closing we remark that although the change of variables between the
covariant and action approaches is rather complicated \cite{BLNPST97b}
our results and those of \cite{GaLaWe98} suggest that, for BPS states, we
may simply identify
\[
\tilde H^{\rm PST}_{mn} = \tfrac14 H^{\rm covariant}_{mnp}v^p,
\]
where $v^p$ is the unit vector in the action formulation \cite{BLNPST97a}.
Perhaps this observation will lead to a better
understanding of the relation between the two approaches.


\begin{thebibliography}{10}

\bibitem{BeSeTe87}
E.~Bergshoeff, E.~Sezgin, and P.~K. Townsend.
\newblock Supermembranes and {E}leven-{D}imensional {S}upergravity.
\newblock {\em Phys. Lett.}, B189:75--78, 1987.

\bibitem{HowSez97}
P.~S. Howe and E.~Sezgin.
\newblock D=11, $p=5$.
\newblock {\em Phys. Lett.}, B394:62, 1997.
\newblock hep-th/9611008.

\bibitem{HoSeWe97a}
P.~S. Howe, E.~Sezgin, and P.~C. West.
\newblock Covariant {F}ield {E}quations of the {M} {T}heory {F}ive-{B}rane.
\newblock {\em Phys. Lett.}, B399:49--59, 1997.
\newblock hep-th/9702008.

\bibitem{PerSch97}
M.~Perry and J.~H. Schwarz.
\newblock Interacting {C}hiral {G}auge {F}ields in {S}ix {D}imensions and
  {B}orn-{I}nfeld {T}heory.
\newblock {\em Nucl. Phys.}, B489:47--64, 1997.
\newblock hep-th/9611065.

\bibitem{AgPaPoSch97}
M.~Aganagic, J.~Park, C.~Popescu, and J.~H. Schwarz.
\newblock World-{V}olume {A}ction of the {M} {T}heory {F}ive-{B}rane.
\newblock {\em Nucl. Phys.}, B496:191--214, 1997.
\newblock hep-th/9701166.

\bibitem{BLNPST97a}
I.~Bandos, K.~Lechner, A.~Nurmagambetov, P.~Pasti, D.~Sorokin, and M.~Tonin.
\newblock Covariant {A}ction for the {S}uper-{F}ive-{B}rane of {M}-{T}heory.
\newblock {\em Phys. Rev. Lett.}, 78:4332--4334, 1997.
\newblock hep-th/9701149.

\bibitem{Wit96}
E.~Witten.
\newblock Five-brane {E}ffective {A}ction {I}n {M}-{T}heory.
\newblock {\em J. Geom. Phys.}, 22:103--133, 1997.
\newblock hep-th/9610234.

\bibitem{LamWes98b}
N.~D. Lambert and P.~C. West.
\newblock Brane {D}ynamics and {F}our-{D}imensional {Q}uantum {F}ield {T}heory.
\newblock In {\em Proceedings of the Trieste Conference on Superfivebranes and
  Physics in 5+1 Dimensions}, April 1998.
\newblock hep-th/9811177.

\bibitem{HoLaWe98}
P.~S. Howe, N.~D. Lambert, and P.~C. West.
\newblock The {S}elf-{D}ual {S}tring {S}oliton.
\newblock {\em Nucl. Phys.}, B515:203--216, 1998.
\newblock hep-th/9710033.

\bibitem{GaLaWe98}
J.~P. Gauntlett, N.~D. Lambert, and P.~C. West.
\newblock Supersymmetric {F}ivebrane {S}olitons.
\newblock preprint QMW-98-40, KCL-TH-98-43, hep-th/9811024, 1998.

\bibitem{HoSeWe97b}
P.~S. Howe, E.~Sezgin, and P.~C. West.
\newblock The {S}ix {D}imensional {S}elf-{D}ual {T}ensor.
\newblock {\em Phys. Lett.}, B400:255--259, 1997.
\newblock hep-th/9702111.

\bibitem{BLNPST97b}
I.~Bandos, K.~Lechner, A.~Nurmagambetov, P.~Pasti, D.~Sorokin, and M.~Tonin.
\newblock On the equivalence of different formulations of the {M} {T}heory
  five--brane.
\newblock {\em Phys. Lett.}, B408:135--141, 1997.
\newblock hep-th/9703127.

\bibitem{GaGoTo98}
J.~P. Gauntlett, J.~Gomis, and P.~K. Townsend.
\newblock {BPS} {B}ounds for {W}orldvolume {B}ranes.
\newblock {\em JHEP}, 01:003, 1998.
\newblock hep-th/9711205.

\bibitem{LamWes98c}
N.~D. Lambert and P.~C. West.
\newblock Monopole {D}ynamics from the {M}-fivebrane.
\newblock preprint KCL-MTH-98-44, hep-th/9811025, 1998.

\end{thebibliography}
\end{document}